# A New Paradigm Integrating the Concepts of Particle Abrasion and Breakage

Priya Tripathi, S.M.ASCE[1]; Seung Jae Lee, Ph.D.[2];
Moochul Shin, Ph.D., A.M.ASCE[3]; and Chang Hoon Lee, Ph.D.[4]

[1]Ph.D. Student, Dept. of Civil & Environmental Engineering, Florida International University, Miami, FL. ORCID: 0000-0003-1743-3219. E-mail: ptrip009@fiu.edu
[2]Assoc. Professor (corresponding author), Dept. of Civil & Environmental Engineering, Florida International University, Miami, FL. ORCID: 0000-0002-2180-3502. E-mail: sjlee@fiu.edu
[3]Assoc. Professor, Dept. of Civil & Environmental Engineering, Western New England University, Springfield, MA. ORCID: 0000-0001-9153-6000. E-mail: moochul.shin@wne.edu
[4]Assistant Professor, Dept. of Civil & Environmental Engineering, Western New England University, Springfield, MA. ORCID: 0000-0001-7209-5203. E-mail: changhoon.lee@wne.edu

## ABSTRACT

This paper introduces a new paradigm that integrates the concepts of particle abrasion and breakage. Both processes can co-occur under loading as soil particles are subjected to friction as well as collisions between particles. Therefore, the significance of this integrating paradigm lies in its ability to address both abrasion and breakage in a single framework. The new paradigm is mapped out in a framework called the *particle geometry space*. The *x*-axis corresponds to the surface-area-to-volume ratio (*A/V*), while the *y*-axis represents volume (*V*). This space facilitates a holistic characterization of the four particle geometry features, i.e., shape (*β*) and size (*D*) as well as surface area (*A*) and volume (*V*). Three distinct paths (abrasion, breakage, and equally-occurring abrasion and breakage processes), three limit lines (breakage line, sphere line, and average shape-conserving line), and five different zones are defined in the particle geometry space. Consequently, this approach enables us to systematically relate the extent of co-occurring abrasion and breakage to the particle geometry evolution.

## INTRODUCTION

Comminution is an important phenomenon in the geology and geotechnical engineering as it greatly influences the macroscopic properties and behavior of geological and granular materials. (Davies and McSaveney 2009; Dufresne and Dunning 2017; Latham et al. 2006; Shen et al. 2022). *Abrasion* and *breakage* are the two primary mechanisms involved in the comminution process. Soil particles experience both friction and collision when subjected to loading, and therefore abrasion and breakage have been observed to co-occur (Bowman et al. 2001; Jin et al. 2022; Qian et al. 2014; Xiao et al. 2022). However, most comminution studies have narrowly focused on either abrasion or breakage (Altuhafi and Coop 2011; Cil and Alshibli 2014; Deiros Quintanilla et al. 2017; Domokos et al. 2009; Einav 2007; Hardin 1985; Harmon et al. 2020; Miller et al. 2014;

Sipos et al. 2021; Wang and Arson 2016; Zhang et al. 2015; Zheng et al. 2019) without considering these processes in a broader framework. Therefore, the breakage studies have reported inconsistent results, which can be attributed to the unaccounted effects of abrasion. For example, Altuhafi and Coop (2011) found that uniaxial compression tests on sand led to an increase of particle angularity. In contrast, Peng et al. (2021) reported a decrease in the shape angularity after particle breakage. Seo and Buscarnera (2022), on the other hand, found "strong correlation between the shape of parent and child particles". It is worth noting that the shape similarity between parent and child particles depends on the parent particles' prior comminution history. For instance, abraded parent particles, thus round shapes, would yield relatively more angular child particles, while freshly crushed parents and their child particles may have relatively similar angularities. Hence, without knowing the history, a simple before-and-after shape comparison may have a limited significance. The objective of this study is to provide a new paradigm that can integrate the concepts of abrasion and breakage. The integrating paradigm is set within the *particle geometry space*. This space enables us to comprehensively present the four particle geometry features, i.e., shape ($\beta$) and size ($D$), surface area ($A$) and volume ($V$) within a single framework. This approach will therefore facilitate a holistic understanding of particle geometry evolution caused by comminution.

**METHODOLOGY**

**Particle Geometry Space**
Lee and colleagues proposed a logarithmic space where the *x*-axis corresponds to particle surface-area-to-volume ratio ($A/V$), and the *y*-axis represents particle volume ($V$) – (Lee et al. 2021, 2022; Su et al. 2020; Tripathi et al. 2023). Hereafter, this space is termed *particle geometry space*. The advantage of using the particle geometry space is the four particle geometry features, i.e., shape ($\beta$), size ($D$), surface area ($A$) and volume ($V$), can be holistically presented. The particle shape index $\beta$ is defined as $A^3/V^2$ in this space which satisfies Eq. (1). The equation can be log-transformed to Eq. (2). Given that *x*- and *y*-axes of the logarithmic particle geometry space represent $A/V$ and $V$ respectively, the $\beta$ corresponds to the *intercept* at $A/V = 1$. If data points representing identical shapes but with varying sizes are plotted in the space, the resulting regression line will exhibit a slope of -3 (e.g., yellow data points as in Figure 1a). The individual particle's shape angularity is presented by $\beta_1$, $\beta_2$, and $\beta_3$, and in this case $\beta_1=\beta_2=\beta_3$. The value of $\beta$ cannot be smaller than $36\pi$ (=113.09) which represents a sphere and gets higher for a more angular shape. For example, $\beta$ is 216 for cube and 374.12 for tetrahedron. It is worth noting that the shape index $\beta$ is equivalent to the inverse of Wadell's true sphericity $S$ as shown in Eq. (3). The size $D$ can be estimated as the diameter of a sphere having the same volume $V$ as the particle as in Eq. (4). Therefore, as in Figure 1, another vertical axis can be adopted to represent $D$.

$$V = (A/V)^{-3} \times \beta \tag{1}$$
$$\log(V) = -3 \times \log(A/V) + \log(\beta) \tag{2}$$
$$(\beta / 36\pi)^{1/3} = ((A^3 / V^2) / 36\pi)^{1/3} = A / A_s = 1 / S \tag{3}$$
$$D = (6V / \pi)^{1/3} \tag{4}$$

Average shape angularity $\beta^*$: The shape angularity of an 'individual particle' could be determined using $\beta$. The 'average' shape angularity can be similarly determined for a 'group of particles' using $\beta^*$. The $\beta^*$ value can be analytically obtained from the arithmetic mean of the $\beta$ values as shown in Eq. (5). Graphically, $\beta^*$ can be also presented as the intercept at $A/V = 1$, obtained from a slope of -3 in the space. For example, $\beta^*$ can be determined from the intercept of the average of the individual data points (indicated by the blue dotted lines in Figure 1).

$$\beta^* = 10^{\overline{\log(\beta)}} = 10^{\overline{\log(V) + 3 \times \log(A/V)}} \tag{5}$$

where $\overline{[...]}$ indicates the arithmetic mean of $[...]$.

Particle shape-size relation $\alpha$: There are three cases that pertain to the relation between particle shape and size, and the slope $\alpha$ of the regression line informs the relation: (a) Case 1: there is no particular relation between shape and size. For example, in case all particles have same angularity regardless of size, we cannot conclude that there is a specific relationship between shape and size. In another scenario, if shapes are random regardless of size, no shape-size relation can be defined. Then, $\alpha$ is equal to -3 ($|\alpha| = 3$) as in Figure 1a; (b) Case 2: there is a tendency that smaller particles (having a smaller $V$) are more angular than larger particles, thus the smaller particles have higher $\beta$ values. In Figure 1b, the particle corresponding to $\beta_1$ is smaller (located at the lowest $V$) than the particle corresponding to $\beta_3$. In this case, the $\alpha$ value for the particle group is greater than -3 ($|\alpha| < 3$); (c) Case 3: there is a tendency that larger particles (having a larger $V$) are more angular. In this case, the $\alpha$ value is smaller than -3 ($|\alpha| > 3$) as in Figure 1c. The Case 3 is uncommon for mineral particles from the reported data (Altuhafi and Baudet 2011; Lee et al. 2022; Tripathi et al. 2023). Therefore, we will focus on Cases 1 and 2 with $|\alpha| \leq 3$ for the remainder of this paper.

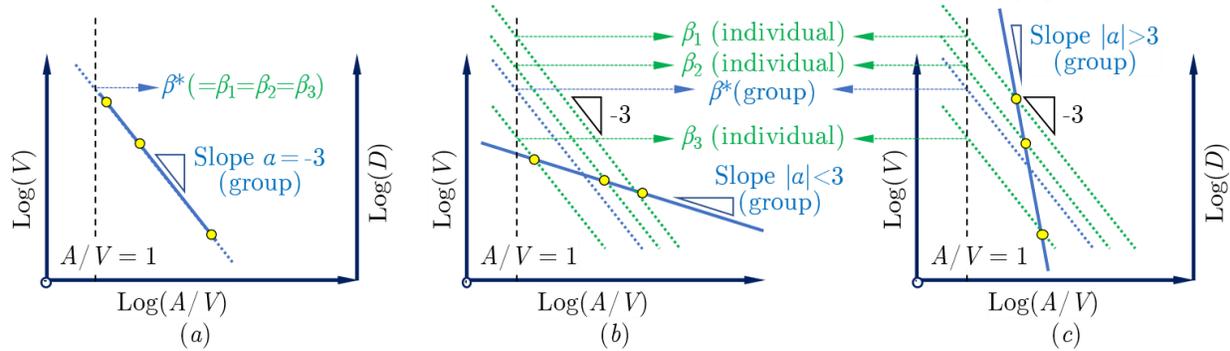

Figure 1. Particle geometry space where $\alpha$ and $\beta^*$ of the regression line inform (i) relation between particle shape and size, and (ii) average shape angularity of a particle group, respectively: (a) Case 1: slope $|\alpha| = 3$ in case there is no particular shape-size relation; (b) Case 2: slope $|\alpha| < 3$ where smaller particles (smaller $V$) are more angular; (c) Case 3: slope $|\alpha| > 3$ where larger particles being more angular.

Lee et al. (2022) interpreted the power regression (i.e., a straight regression line in the logarithmic particle geometry space) as a *phenotypic trait of particle geometries*, influenced by the shared geological origin and loading histories within the group. Each group therefore displays a distinct power regression line representing this trait. The phenotypic trait is a concept encompassing both shape and size. Similar to a race of people sharing a common genetic origin despite different appearances, a family of mineral particles share a phenotypic trait due to a common geologic origin despite varying shapes and sizes. Therefore, shapes can differ while the data points collectively form a regression line, e.g., Case 2 (Figure 1). As the phenotypic trait is inherited across generations in a race of people, we hypothesize that this power regression continues to hold true in the process of comminution for a family of particles.

**Particle Geometry Evolution**

**Three axioms.** This study employs the particle geometry space to map the particle geometry evolution caused by both abrasion and breakage. The mapping is built on the following three axioms (ax. 1 – 3) that are considered as true:

- ax.1 - comminution reduces particle size (regardless of abrasion or breakage).
- ax.2 - abrasion results in the rounding of particle shapes.
- ax.3 - breakage produces particles with angular shapes.

The three axioms are schematically shown in Figure 2a, where $\{P_O\}$ represents an initial set of particles. With ax.1, $\{P_O\}$ always moves down to a lower $V$ due to a decrease in the particle size. With ax.2, $\{P_O\}$ moves towards the left as abrasion leads to a rounding of particle shape, accompanied by a decrease in $\beta^*$. In contrast, with ax.3, $\{P_O\}$ moves to the right, increasing $\beta^*$.

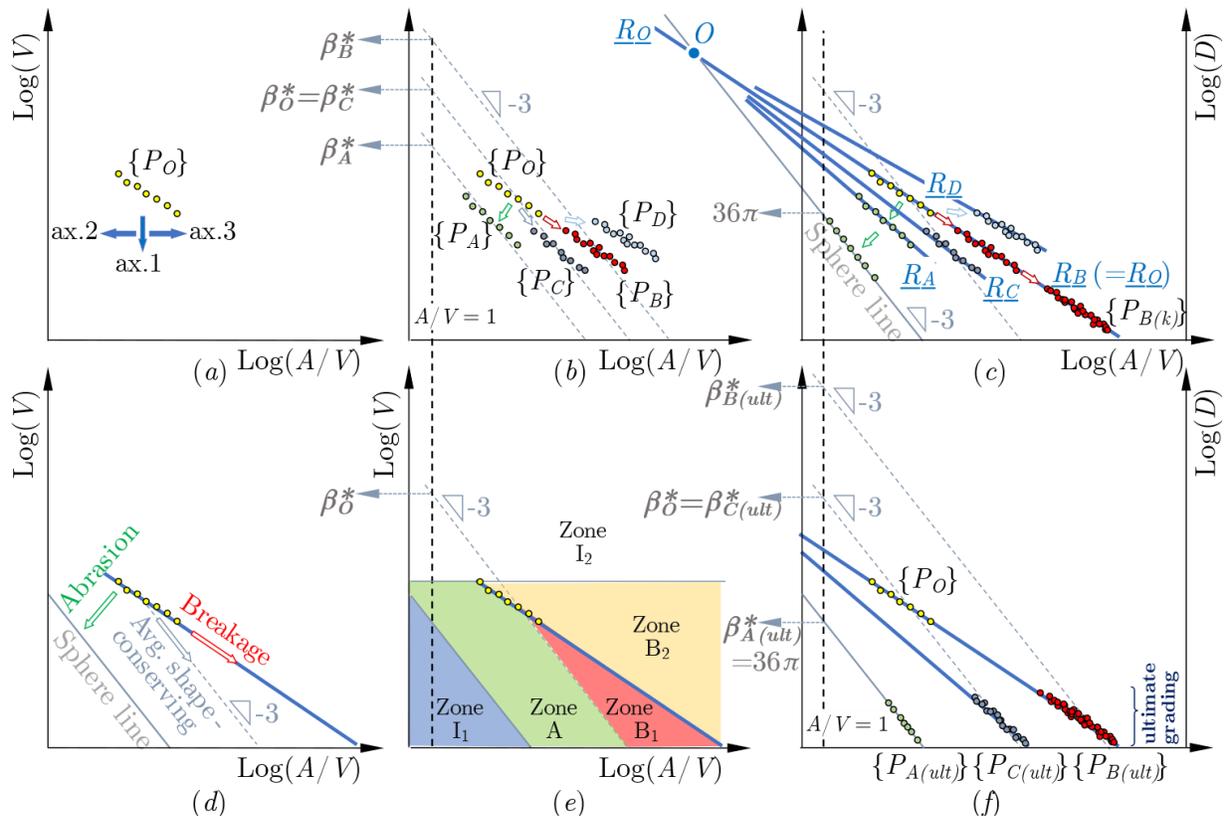

Figure 2. Evolution of particle geometry due to abrasion and breakage mapped on the particle geometry space.

**Change in the intercept β\* during comminution.** An illustration is shown in Figure 2b, where $\{P_A\}$ and $\{P_B\}$ represent the particle sets after abrasion and breakage, respectively. Abrasion, under ax.1 & 2, transforms $\{P_O\}$ to $\{P_A\}$, shifting data points lower left. Breakage, under ax.1 & 3, turns $\{P_O\}$ into $\{P_B\}$. It is worth noting that the number of data points will change as $\{P_O\} \rightarrow \{P_B\}$, producing 'child' particles, unlike $\{P_O\} \rightarrow \{P_A\}$ without major fragmentations. The $\beta^*_O$, $\beta^*_A$, and $\beta^*_B$ represent the average angularity of $\{P_O\}$, $\{P_A\}$, and $\{P_B\}$, respectively, with $\beta^*_A < \beta^*_O < \beta^*_B$ indicating the angularity decrease or increase caused by the comminution. In a 'special' comminution process where abrasion and breakage occur equally, the initial shape angularity $\beta^*_O$ would be conserved. Then, the resulting particle set $\{P_C\}$ will be on a line with slope of -3 from $\{P_O\}$. The $\beta^*$ is determined at the $A/V = 1$ intercept on this line with slope of -3, making it the only possible path for $\beta^*_O = \beta^*_C$. The line with slope of -3 is termed the *average shape-conserving line*.
<u>Breakage of initially abraded particles</u>: If $\{P_O\}$ has already experienced significant abrasion, breakage will cause a more dramatic shape change. The resulting particles, $\{P_D\}$, will appear far-right of $\{P_O\}$, as shown in Figure 2b, with much higher $\beta^*$ values than $\beta^*_O$.

**Change in the power regression's slope α during comminution.**
<u>Abrasion</u>: The regression line with initially $|\alpha| < 3$ gradually converges towards $|\alpha| = 3$, as angular particles typically undergo abrasion faster than round particles (Janoo 1998; Krumbein 1941), making all shapes more similar to one another. In an extreme abrasion, the data points cluster near the *sphere line* ($|\alpha| = 3$ and $\beta^* = 36\pi$). Figure 2c illustrates the concept. The <u>$R_O$</u> and <u>$R_A$</u> represent the regression lines corresponding to $\{P_O\}$ and $\{P_A\}$, respectively. The <u>$R_O$</u> starts with $|\alpha| < 3$, and the slope gets steeper with abrasion, thus <u>$R_A$</u> is closer to $|\alpha| = 3$.
<u>Breakage</u>: The child (broken) particles inherit parents' traits, yielding the identical regression line. See Figure 2c. <u>$R_B$</u> is the regression line that corresponds to $\{P_B\}$, realizing <u>$R_B$</u> = <u>$R_O$</u>. The set $\{P_{B(k)}\}$ denotes the resulting particles after *k*-th continuous breakage. Further breakage maintains the line such that <u>$R_{B(k)}$</u> = <u>$R_O$</u> due to the inherited trait. In this study, this line is termed the *breakage line*.
<u>Equally-occurring abrasion and breakage</u>: While $\{P_O\} \rightarrow \{P_C\}$ occurs along a line with slope of -3 (keeping $\beta^*$ constant), the slope $|\alpha|$ of <u>$R_C$</u> (regression line corresponding to $\{P_C\}$) becomes steeper than the slope $|\alpha|$ of <u>$R_O$</u> due to the effect of abrasion making all shapes similar to one another.
<u>Geometric origin of comminution</u>: The point *O*, where the *breakage line* and *sphere line* intersect, is termed the *geometric origin of comminution*. It serves as the key reference for the slope $|\alpha|$ during comminution processes, with the regression lines pivoting around it.
<u>Breakage of initially abraded particles</u>: If $\{P_O\}$ has experienced significant abrasion in the prior history, the slope $|\alpha|$ of <u>$R_O$</u> will be already high. It will be close to 3, if $\{P_O\}$ was extremely abraded. Upon breakage, a new breakage line <u>$R_D$</u> will emerge, and the slope will exhibit what could have been obtained from the freshly crushed parent particles prior to undergoing abrasion.

**Five zones in the *A/V* and *V* space.** With consideration of three distinct paths (abrasion, breakage, and equally-occurring abrasion and breakage processes) and three limits (sphere line, breakage line, and average shape-conserving line) as in Figure 2d, five zones can be defined as in Figure 2e:

(1) Zone A on the left of the average shape-conserving line informs *more abrasion* because the data points in this zone realize a lower $\beta^*$ than $\beta\bar{o}$; (2) Zone $B_1$ is located on the right side of the average shape-conserving line and informs *more breakage*. The data points in this zone realize higher $\beta^*$ values than $\beta\bar{o}$; (3) Zone $B_2$ is on the right side of the initial regression line $\underline{R_O}$. If the particles data show up in this zone after breakage, it informs the parent particles have undergone some significant abrasion prior to breakage; (4) Zone $I_1$ is underneath the sphere line. This is an 'impossible zone' where data cannot exist because there is no rounder shape than a sphere; (5) Zone $I_2$ is located above $\{P_O\}$. This is another impossible zone because particles do not get larger after comminution.

**Ultimate shape angularity.** The shape angularity in extreme comminution will be determined at the ultimate grading, where no further changes in particle size distribution are expected (Altuhafi and Coop 2011; Einav 2007). In Figure 2f, the particle sets after the ultimate grading are labeled as $\{P_{A(ult)}\}$, $\{P_{B(ult)}\}$, and $\{P_{C(ult)}\}$ resulting from extreme abrasion, breakage, and an equally-occurring abrasion and breakage process, respectively. The data points sit at the lowest possible $V$ from the ultimate grading, with the average angularities evaluated as $\beta^*_{A(ult)} < \beta^*_{C(ult)} (= \beta\bar{o}) < \beta^*_{B(ult)}$. In an extreme abrasion, $\{P_{A(ult)}\}$ becomes spherical, aligning data points on the sphere line.

**EXPERIMENTAL VALIDATION**

Comminution data obtained from experiments are mapped onto the particle geometry space to demonstrate the paths $\{P_O\}\rightarrow\{P_A\}$ and $\{P_O\}\rightarrow\{P_B\}$. The abrasion test data reported by Paixão and Fortunato (2021) are utilized for $\{P_O\}\rightarrow\{P_A\}$. A series of drop hammer test is conducted for $\{P_O\}\rightarrow\{P_B\}$ to induce breakage without abrasion.

**Abrasion**
Paixão and Fortunato (2021) conducted micro-Deval abrasion tests on 30 granite particles, and 3D scanned those particles at 0, 2000, and 14000 revolutions. Figure 3 plots this geometry data, initially showing $|\alpha|$ at 1.8 ($< 3$), which suggests smaller particles are more angular. After 14000 revolutions, $|\alpha|$ increases to 1.92 due to more abrasion on angular particles, resulting in shapes relatively uniform than the initial condition at 0 revolution. The data points shift lower left, decreasing $\beta^*$ value from 260.98 to 226.72, and approaching the sphere line with $\beta^* = 36\pi$. This data demonstrates the $\{P_O\}\rightarrow\{P_A\}$ path.

**Breakage**
A sample of 10 particles collected from a batch of crushed granite particles is used for this study. The batch was obtained from a quarry located in Richmond, Virginia. Figure 4 shows the 10 Virginia granite (VG) particles. A drop hammer test is conducted on each particle to induce breakage without abrasion, dropping a 5.5 lbs. Proctor hammer from 2.5 inches high. Broken particles larger than #10 sieve size (2 mm) are sifted and shown in Figure 4. The loss of finer particles is about 12%. A Polyga C504 3D scanner is used to capture the geometry before and after the test, from which particle surface area (*A*) and volume (*V*) are obtained. Figure 5 plots the geometry changes. The child particles inherit the phenotypic trait of the parent particles, thus both data sets share the same regression line (Figure 5a), while the average angularity β* increases from 285.90 to 389.11 (Figure 5b). This experiment demonstrates the $\{P_O\} \rightarrow \{P_B\}$ path.

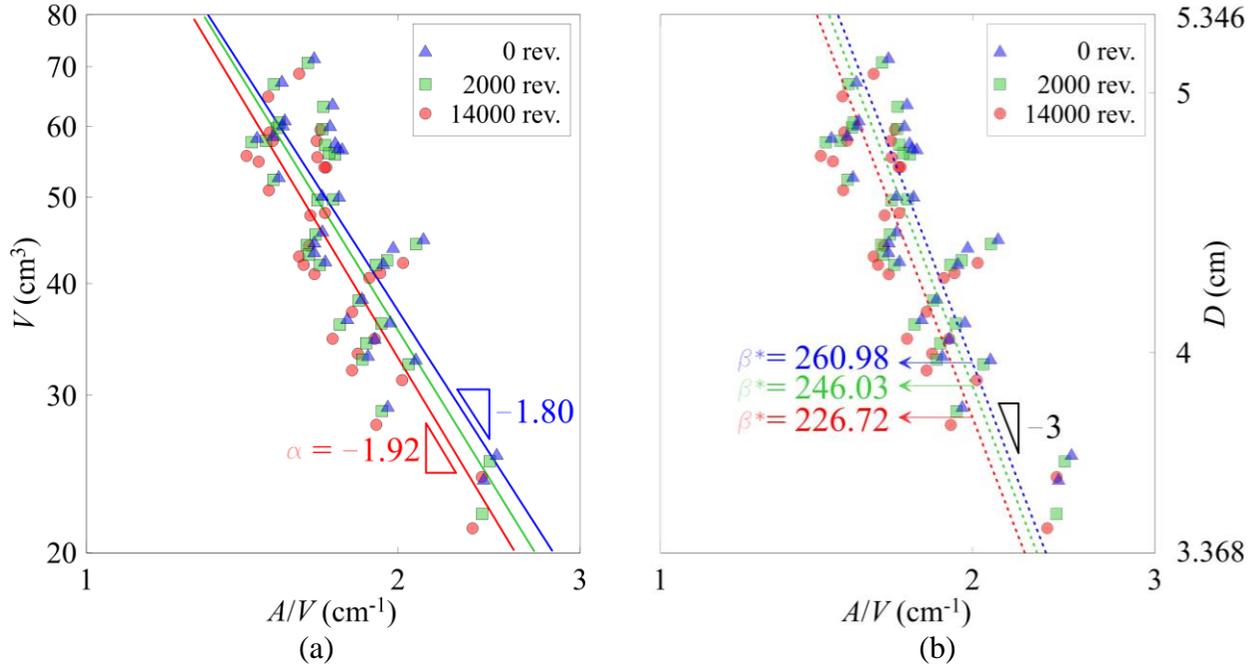

Figure 3. Particle geometry evolution by abrasion: (a) The power regression's slope |*α*| increases; (b) The intercept *β** decreases, suggesting the data points approaching the sphere line.

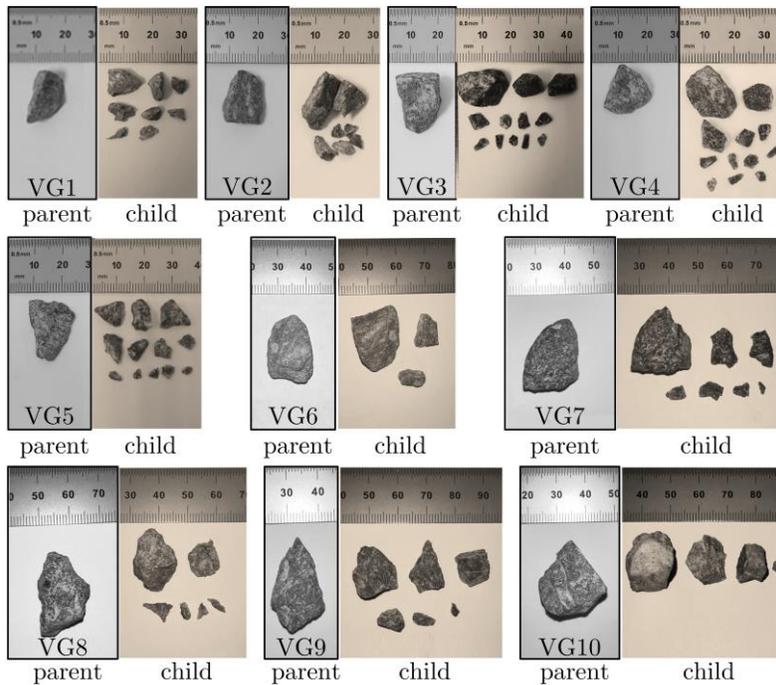

Figure 4. Virginia granite (VG) Particles before and after breakage

Tripathi et al. (2023)  – 7 –

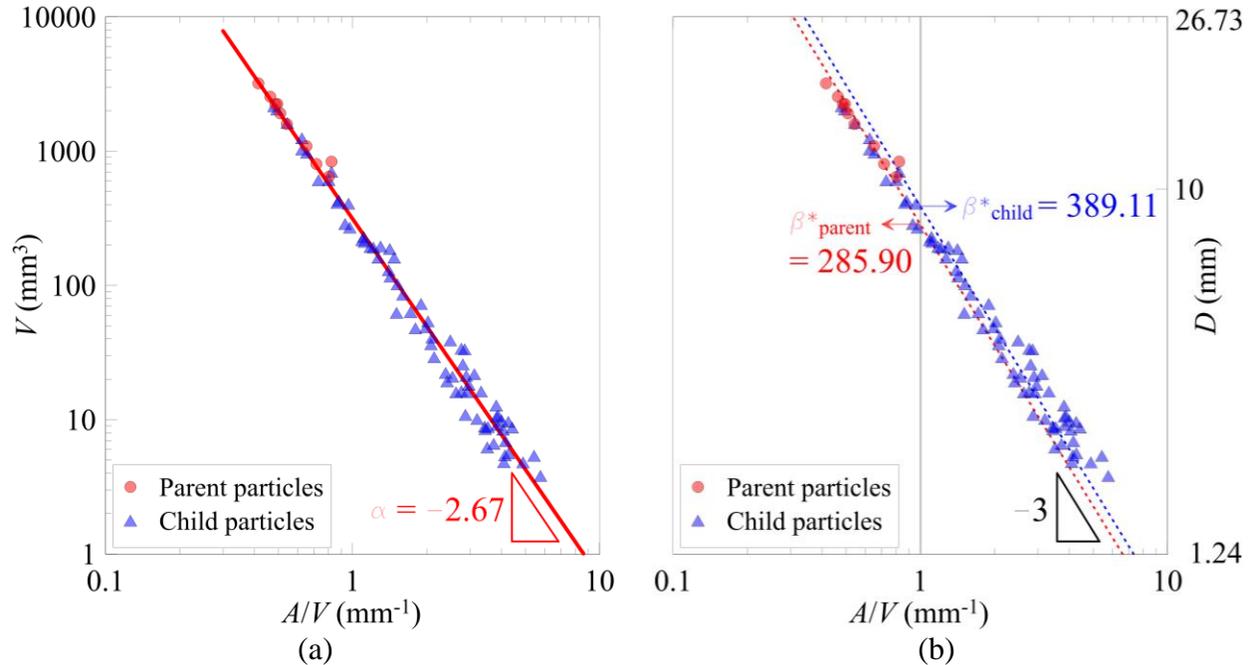

Figure 5. Particle geometry evolution by breakage: (a) The child particles inherit the phenotypic trait of parent particles, sharing the same regression line; (b) The intercept $\beta^*$ increases.

## CONCLUDING REMARKS

This study presents a new paradigm for understanding the particle geometry evolution due to abrasion and breakage. The particle geometry space is used to integrate the concepts of abrasion and breakage. This approach allows for mapping the geometry evolution within a single framework, effectively capturing the changes of shape ($\beta$), size ($D$), surface area ($A$), and volume ($V$). Three distinct paths, three limit lines, and five in-between zones are defined in the space to systematically map the particle geometry evolution resulting from both abrasion and breakage. The proposed approach will be able to serve as a new conceptual foundation, enabling a comprehensive and coherent understanding of the complex comminution processes.

## DATA AVAILABILITY

The dataset of digital particles generated from this study is available in the NSF DesignSafe-CI repository (Tripathi et al. 2023). Please see the Virginia granite group B particles in the dataset.

## ACKNOWLEDGEMENTS

This work is partly sponsored by the US National Science Foundation (NSF) under the awards CMMI #1938431 and #1938285. The support is greatly appreciated. The authors extend their appreciation to Dr. Eric Koehler in Titan America LLC for providing the Virginia granite particles used in the breakage study.